\documentstyle[aps]{revtex}
\def\gtrsim
{\raise2pt\hbox
{$\displaystyle{}\mathrel{\mathop>_{\raise1pt\hbox{$\sim$}}}{}$}} 
\def\lesssim
{\raise2pt\hbox
{$\displaystyle{}\mathrel{\mathop<_{\raise1pt\hbox{$\sim$}}}{}$}}
\begin{document}
\title{
Domain Wall Resistance based on Landauer's Formula} 
\author{Gen {\sc Tatara}\\
Graduate School of Science, Osaka University, Toyonaka, Osaka 560-0043}
\maketitle
\begin{abstract}
The scattering of the electron by a domain wall in a nano-wire is 
calculated perturbatively to the lowest order. The resistance 
is calculated by use of Landauer's formula. 
The result is shown to agree with the result of the linear response theory
if the equilibrium is assumed in the four-terminal case.
\end{abstract}
\sloppy

\section{Introduction}
The electronic transport properties in nano-size ferromagnetic structures 
has 
been drawing growing interest recently. Of most interest would be 
the magnetoresistance effect, namely, the change of the resistance by 
application of a magnetic field. The magnetoresistance in nano-structures 
is due to the rearrangement of the 
magnetization, namely the creation and motion of domain walls, driven 
by the external field.
Recently high-sensitivity 
measurements on magnetic nano-structures such as wires and 
contacts has become possible and intensive experimental studies of the 
magnetoresistance due to a single domain wall have 
begun\cite{Hong96,Otani97,Ruediger98}
\cite{Ruediger98c,Garcia99,Ono99,TZMG99,Komori99,Zhao99}.
These recent experiments have evoked a renewal of interest in
theories of the scattering of the electron by a domain 
wall\cite{TZMG99,Yamanaka96,TF97,Zhang97,Geller98,vanHoof99,Brataas99}
\cite{Imamura00,Nakanishi99}, 
although the first theoretical study of this problem goes back to more 
than 25 years ago\cite{Cabrera74}.

The scattering mechanism is due to the non-adiabatic deviation 
of the electron spin from the local magnetization as the electron 
goes through the wall. In 
the case of a wall with thickness $\lambda \gg k_{F}^{-1}$ 
($k_{F}$ being the Fermi wave length), the electron spin can 
adiabatically follow the spatial rotation of the magnetization, 
resulting in a negligiblly small reflection. 
This is the reason the wall contributes very little in the 
magnetoresistance in the 3$d$ metals in the bulk\cite{Cabrera74}.

The situation changes in nano-magnets, where the profile of the wall 
can be different from that in the bulk magnets. 
The extreme case would be an atomic size 
contact. There the wall is believed to be constrained in a small contact 
region of an atomic scale due to a rapid change of the contact 
shape\cite{Bruno99}, and thus the adiabaticity does 
not hold resulting in a strong scattering\cite{vanHoof99,TZMG99}.
Indeed an experiment on Ni contact revealed a magnetoresistance of 
about 300\% at room temperature\cite{Garcia99}, which is 
significantly large compared with other giant magnetoresistive 
materials such as multilayers. 
The result has been interpreted as due to a strong electron scattering by 
a domain wall, which is trapped in the narrow contact region in the 
absence of the magnetic field, and which is expelled by applying a 
magnetic field\cite{TZMG99}. 

In this paper the domain wall resistance in nano-wires or contacts
is calculated by use of Landauer's formula.
The advantage of this formalism is that we can see how the density 
of states of the sample and the lead appears in the resistance, which 
is not possible in he linear response calculation as carried out in 
ref. \cite{TZMG99}.
The system is treated as in one dimension and the calculation is at 
zero temperature.
The transmission coefficients are calculated in section 2, and the 
resistance is calculated in the case of four- and two-terminals in 
section 3. The result is shown to agree with the result of the Mori 
formula of the linear response theory in the case of the equilibrium
flow in a four-terminal case.

Throughout the paper we consider only channels which contribute to the 
transport and so
do not discuss the quantized behavior of the conductance.

The phase of the transmission coefficient contains important 
information on the stability of the electron system. 
This has been considered elsewhere\cite{TT00}.

\section{Scattering Properties}
We consider the exchange 
interaction between the electron spin and the local spin.
The Hamiltonian is

\begin{equation}
\label{1}H=\sum\limits_{k\sigma }\epsilon _k c_{k\sigma }^{\dagger
}c_{k\sigma }-g\int dx{\bf S}(x)(c^{\dagger }\mbox{\boldmath$\sigma$}c),
\end{equation}

where $\epsilon _k\equiv \hbar ^2k^2/2m-\epsilon _F$ ($\epsilon _F$ being
the Fermi energy). The spin index is denoted by $\sigma =\pm $ and 
$\mbox{\boldmath$\sigma$}$ is the Pauli matrix
(spin indices are suppressed in the second term).
The local spin ${\bf S}$ has a spatial dependence of a domain wall.
In terms of a polar coordinate $\left( S_z\equiv S\cos \theta \right) $ the
wall is represented as $\cos \theta\rightarrow 1$ for $x\ll -\lambda$ 
and $\cos \theta\rightarrow-1$ for $x\gg\lambda$,
$\lambda $ being the thickness of the wall. 
We consider two typical profiles of the wall;
\begin{eqnarray}
	 \mbox{\rm bulk wall}&: & \cos\theta=\tanh(x/\lambda) 
		\label{bulkwall} \\
	        \mbox{\rm linear wall} &: & \cos\theta= \left\{
	             \begin{array}{cc} 1  & (x>\lambda) \\
	                      x/\lambda & (-\lambda \leq x \leq \lambda) \\
	                      -1 & (x<-\lambda) \end{array}\right. , 
	\label{DWprofiles}
\end{eqnarray}
where the first case corresponds to the wall determined by the 
uniaxial anisotropy energy, which would be the case of bulk magnets, 
and the second linear profile would be realized in small contacts.

To proceed we 
carry out a local gauge transformation in the electron spin space so that
the $z$-axis is chosen to be along the local direction of spin 
$\overrightarrow{S}$\cite{TF94}. The transformation is written as
$a_\sigma =(Uc)_\sigma$, where $U$ is a position-dependent matrix defined 
as
$U\equiv \cos \left(\frac \theta 2 \right) \sigma_z+ 
\sin \left(\frac \theta 2 \right) \sigma_y$ and
the electron in the new frame is denoted by $a_\sigma $. The
Hamiltonian is transformed to be
$
H=H_{0}+H_{\rm int}
$
where 
$H_{0}\equiv \sum\limits_{k\sigma }\epsilon _{k\sigma }a_{k\sigma }^{\dagger
}a_{k\sigma }$ and 
$\epsilon _{k\sigma }\equiv \epsilon _k-\sigma \Delta
$is the energy of uniformly polarised electron with the exchange splitting 
$\Delta \equiv g\left| {\bf S}\right|$. The term $H_{\rm int}$ describes the
interaction between the electron and the wall and is given by 
\cite{TF97}

\begin{equation}
\label{DWint}H_{\rm int}\equiv \frac{\hbar ^2}{2m}\frac 1L\sum\limits_{kq}\left[
-\left( k+\frac q2\right) A_qa_{k+q}^{\dagger }\sigma _xa_k+\frac 1{4L}%
\sum\limits_pA_pA_{-p+q}a_{k+q}^{\dagger }a_k\right] ,
\end{equation}
where

\begin{equation}
\label{6}A_q\equiv \int dxe^{-iqx}\nabla \theta,
\end{equation}
is a domain-wall form factor.
Note that $\nabla \theta\rightarrow 0$ for $|x|\rightarrow \infty$ 
and so the interaction is localized near the wall. 
The spin-conserving process in the rotated frame as the electron goes 
through the wall corresponds to the adiabatic process, where the 
electron spin follows the rotation of the magnetization, and the spin 
flip corresponds to the deviation from the adiabaticity.

The interaction $H_{\rm int}$ leads to a scattering of the electron 
and thus contributes to a resistivity due to the 
wall as has been discussed before based on a linear response 
theory\cite{TF97,Brataas99,TZMG99}.  
Here we take a different approach of 
calculating the $T$-matrix perturbatively with respect to $H_{\rm int}$.
The solution of the Schr\"odinger equation with the energy of $E$, 
which we denote by $|\Psi_{E}>$ (two component 
vector corresponding to two spin state), is given as
 \begin{equation}
 	|\Psi_{E}>=|\Psi_{0}>+G^0_{E}T_{E}|\Psi_{0}>,
 	\label{Psidef}
 \end{equation}
where $|\Psi_{0}>$ is the incoming wave, $G^0_{E}\equiv 
\frac{-1}{H_{0}-E-i0}$ is the free Green function with energy $E$ 
and $T$-operator is 
defined as $T_{E}\equiv H_{\rm int} \frac{1}{1-G^0_{E}H_{\rm int}}$. The small 
imaginary part in  $G^0_{E}$ is to choose the correct boundary 
condition of outgoing wave. 
We consider the incoming wave with energy $E$ and spin $\sigma$. 
They are explicitly written as
\begin{eqnarray}
	<x|\Psi^\uparrow_{0}> &=& 
	\left(\begin{array}{c}  e^{ik_{+}x} \\
	                         0   \end{array}\right), 
	<x|\Psi^\downarrow_{0}> = 
	\left(\begin{array}{c}   0 \\
	                         e^{ik_{-}x}   \end{array}\right),
	\label{psi0}
\end{eqnarray}
where $k_{\sigma}\equiv \sqrt{2m(E+\sigma\Delta)}$ is the wave vector 
of the spin $\sigma$ component.
The spin $\sigma'$ component of the full wave function for the 
incident wave $|\Psi^\sigma_0>$ is then obtained from eq. 
(\ref{Psidef}) as
\begin{equation}
	\Psi_{E}^{\sigma'\sigma}(x)=\delta_{\sigma',\sigma}e^{ik_{\sigma}x}
	+\int dx' G^0_{E\sigma'}(x-x') 
	\sum_{k'}e^{ik'x'}T_{k'\sigma',k_\sigma \sigma},
	\label{Psi1}
\end{equation}
where $G^0_{E\sigma'}(x-x') \equiv <x\sigma'|G^0_{E}|x'\sigma'> 
=-i\frac{m}{|k_{\sigma'}|} e^{ik_{\sigma'}|x-x'|}$ and
$T_{k'\sigma',k_\sigma \sigma}\equiv <k'\sigma'|T_{E}|k_{\sigma}\sigma>$.

The transmission and reflection coefficients are defined from the 
asymptotic behavior as
\begin{eqnarray}
	\Psi_{E}^{\sigma'\sigma}(x) &\rightarrow & \left\{
	\begin{array}{cc}
   \delta_{\sigma'\sigma}e^{ik_{\sigma}x}+r_{\sigma'\sigma}e^{-ik_{\sigma'}x} 
   &  (x\rightarrow -\infty) \\
     t_{\sigma'\sigma}e^{ik_{\sigma'}x} 
   &  (x\rightarrow +\infty)  \end{array} \right.   .
   	\label{rtdef}
\end{eqnarray}
The result is
\begin{eqnarray}
	t_{\sigma'\sigma}(k_{\sigma}) & = & 
	\delta_{\sigma'\sigma}-i\frac{mL}{|k_{\sigma'}|} 
	T_{k_{\sigma'}\sigma',k_{\sigma}\sigma} \nonumber\\
	r _{\sigma'\sigma}(k_{\sigma}) & = & -i\frac{mL}{|k_{\sigma'}|} 
	T_{-k_{\sigma'}\sigma',k_{\sigma}\sigma} .
	\label{tr}
\end{eqnarray}
The diagonal (in spin) component of the 
$T$-matrix for the incoming wave vector of $k_{\sigma}$
is calculated as
\begin{eqnarray}
	T_{k_{\sigma}\sigma,k_{\sigma}\sigma} &=& 
	<k_\sigma,\sigma|H_{\rm int}|k_\sigma,\sigma> 
	+\sum_{k'}<k_\sigma,\sigma|H_{\rm int}|k',-\sigma> 
<k',-\sigma|G_{E}^0|k',-\sigma> <k',-\sigma|H_{\rm 
int}|k_\sigma,\sigma>\nonumber\\
&& +O(A^4) \nonumber\\
  &=& \frac{1}{8mL^2}\sum_{q} |A_{q}|^2 \left[1-
	\frac{(2k_{\sigma}+q)^2} 
	{(q+k_{\sigma})^2-k_{-\sigma}^2-i0}\right]+O(A^4).
	\label{Tdiag}
\end{eqnarray}
Off-diagonal components are given by
\begin{equation}
	T_{\pm k_{-\sigma},-\sigma,k_{\sigma}\sigma} = 
\frac{(k_{\sigma}\pm k_{-\sigma})}{4mL} A_{\pm k_{-\sigma}-k_{\sigma}}
+O(A^3).
	\label{Toffdiag}
\end{equation}
The result of the transmission and reflection coefficients are 
\begin{eqnarray}
	t_{\sigma\sigma}(k_{\sigma}) &=& 
	1+i\frac{1}{8k_{\sigma}L}\sum_{q} |A_{q}|^2 
	\frac{3k_{\sigma}^2+k_{-\sigma}^2+2k_{\sigma} q} 
	{(q+k_{\sigma})^2-k_{-\sigma}^2-i0} \nonumber\\
	t_{-\sigma,\sigma}(k_{\sigma}) &=& 
	 -i\frac{(k_{\sigma}+k_{-\sigma})}{4k_{-\sigma}} 
	 A_{-k_{\sigma}+k_{-\sigma}}  \nonumber\\
	r_{-\sigma,\sigma}(k_{\sigma}) &=& 
	 -i\frac{(k_{\sigma}-k_{-\sigma})}{4k_{-\sigma}} 
	 A_{-k_{\sigma}-k_{-\sigma}}  \nonumber\\
	 r_{\sigma,\sigma}(k_{\sigma}) &=&  O(A^2)	 .
	\label{t}
\end{eqnarray}
The scattering probabilities
are defined as $T_{\sigma'\sigma}\equiv|t_{\sigma'\sigma}|^2$ and  
$R_{\sigma'\sigma}\equiv|r_{\sigma'\sigma}|^2$.
These probabilities for the electron at Fermi level (i.e., 
$k_{\sigma}=k_{F\sigma}$) are obtained to the 
second order in $A$ as
\begin{eqnarray}
	T_{\sigma\sigma} & = & 1-\frac{1}{4(1-\zeta^2) }
	(\zeta^2|A_{+}|^2+|A_{-}|^2)
	\nonumber  \\
	T_{-\sigma,\sigma} & = & \frac{1}{4(1-\sigma\zeta)^2 }
	|A_{-}|^2
    \nonumber  \\
	R_{-\sigma,\sigma}& = & \frac{1}{4(1-\sigma\zeta)^2} |A_{+}|^2,
	\label{TRresults}
\end{eqnarray}
and $R_{\sigma\sigma}=O(A^4)$, where $A_{+}\equiv 
A_{k_{F+}+k_{F-}}(=A_{2k_{F}})$ and 
$A_{-}\equiv A_{k_{F+}-k_{F-}}(=A_{2k_{F}\zeta})$ 
describes a backward and forward 
scattering with spin flip, respectively ( 
$k_{F}\equiv(k_{F+}+k_{F-})/2$ and 
$\zeta\equiv (k_{F+}-k_{F-})/(k_{F+}+k_{F-})$). 
The conservation of the current,
\begin{equation}
k_{F\sigma}=k_{F\sigma}T_{\sigma\sigma}+k_{F,-\sigma}T_{-\sigma,\sigma} 
+k_{F,-\sigma}R_{-\sigma,\sigma},
\end{equation}
is satisfied.

\section{Wall Resistance}
\begin{figure}
%
\caption{The system coupled with four terminals. $L_{1}$ and $L_{2}$ 
are leads with chemical potential $\mu_{1}\equiv \epsilon_{F}+eV_0$ and 
$\mu_{2}\equiv \epsilon_{F}$, respectively, which are 
coupled with an infinitely large current source.
$T_{A}$ and $T_{B}$ are the terminals to measure voltage whose 
chemical potential is denoted by $\mu_{A}$ and $\mu_{B}$, respectively. 
The voltage of the sample is given by $V=(\mu_{A}-\mu_{B})/e$.
\label{FIGterminal}}
\caption{Schematic picture of an equilibrium flow in the scattering 
of particles with two components with different velocities. If some 
portion of the incoming particle with velocity $\hbar k_{F+}/m$ is 
scattered to have a smaller velocity of $\hbar k_{F-}/m$, the density
of these outgoing particle becomes higher. 
\label{FIGdensity}}
\caption{
Resistances by the wall for two different profiles of the wall (bulk 
wall given by
eq. (\ref{bulkwall}) (upper graph) and linear wall, 
(\ref{DWprofiles}) (lower graph)) 
plotted as functions of $\lambda$ for $\zeta=0.3$ and $0.8$.
Solid and dashed lines corresponds to the four-terminal case ($R_{\rm w}$)
 and two-terminal case ($R_{\rm w}^{\rm (2t)}$), respectively.
\label{FIGR}
}\end{figure}
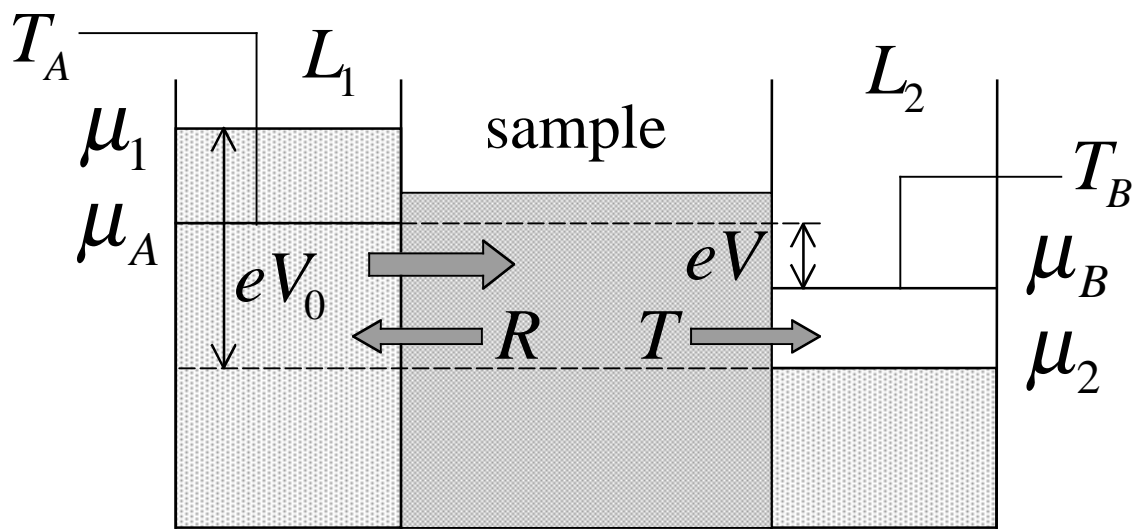
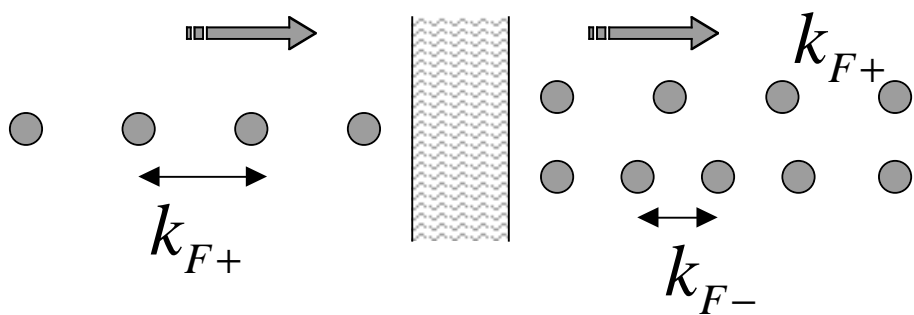
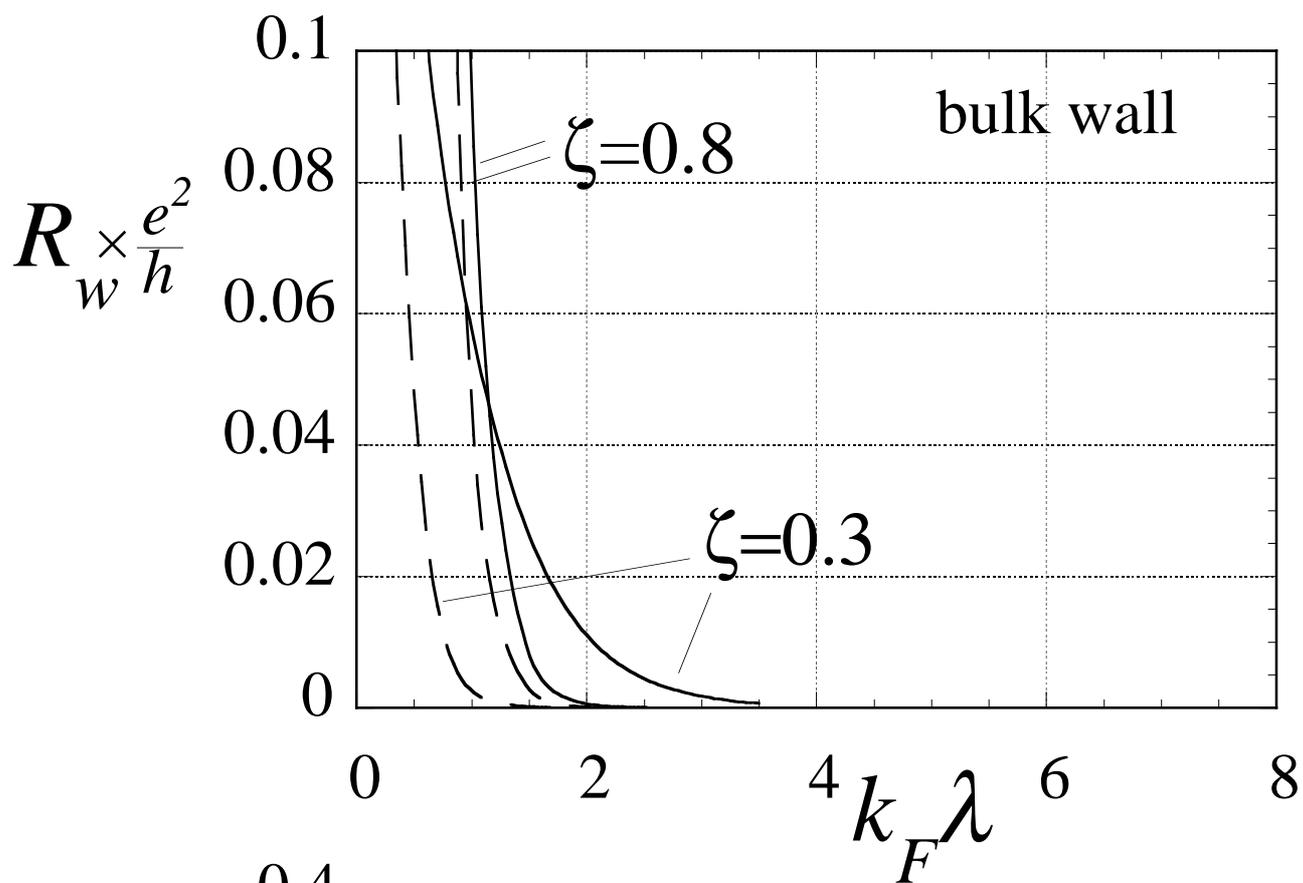
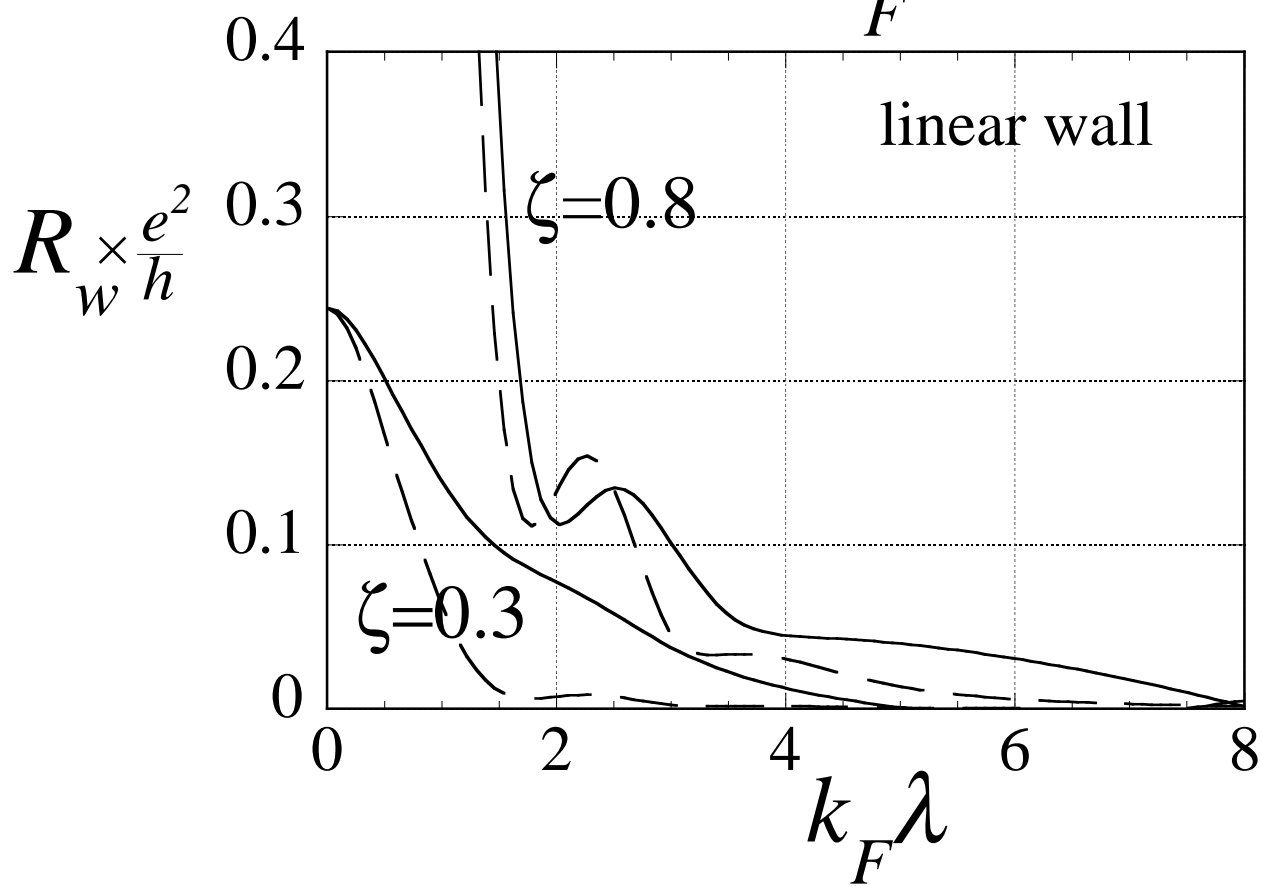

To relate these quantities to resistance, we consider a
four-terminal case where the system 
is coupled with two leads, $L_1$ and $L_2$ and two 
terminals, $T_{A}$ and $T_{B}$ (Fig. 
\ref{FIGterminal}). 
(The gauge transformation does not affect the resistance, 
since the shift of the 
current due to the transformation vanishes far away from the 
wall.)
$L_{1}$ and $L_2$ are connected with a reserver (current source) and
$T_{A}$ and $T_{B}$ are to measure the voltage difference across the sample.
The argument goes similarly to the non-magnetic case 
described in refs. \cite{Buttiker85,Imry97} but taking account of 
spin dependence. 
The chemical potential of $L_{1}$ and $L_2$ are fixed to be 
$\mu_{1}\equiv\epsilon_{F}+eV_{0}$ and $\mu_{2}\equiv\epsilon_{F}$ ($V_{0}$ is 
the applied voltage), 
respectively. The chemical potential of $T_{A}$ and $T_{B}$, 
$\mu_{A}$ and $\mu_{B}$ are determined below by the condition of vanishing 
current flow through each terminal.
The true voltage difference across the sample is different from 
$V_{0}$ and is given as $V\equiv(\mu_{A}-\mu_{B})/e$.
The resistance of the system is written as $R_{\rm w}\equiv 
V/I$, $I$ being the current through the sample. 

We first consider the case where the leads are ferromagnets with 
density of states for the spin state $\sigma$ given by $N_{\sigma}$. 
The case of non-magnetic leads is discussed at the end of this section.
The number of the electron with spin $\sigma$ injected from $L_{1}$ is 
$N_{\sigma}eV_{0}$, and thus the electric current is 
written by use of transmission probability as
\begin{equation}
	I=eV\sum_{\sigma,\sigma'}N_{\sigma}\frac{e\hbar}{mL}k_{F\sigma'}
	 T_{\sigma'\sigma}.
	\label{Ieq}
\end{equation}
The voltage $\mu_{A}$ and $\mu_{B}$ are determined by the balance 
between the terminal and lead. 
In $L_{1}$, the density (per energy) of spin $\sigma$ electron is 
written as
\begin{equation}
	n_{1\sigma}=\frac{1}{2}
	   (N_{\sigma}(1+R_{\sigma\sigma})+N_{-\sigma}R_{\sigma,-\sigma}).
	\label{n1}
\end{equation}
Here a factor of $1/2$ comes from choosing right moving 
electron. The density at the other lead is given by
\begin{equation}
	n_{2\sigma}=\frac{1}{2}
	   (N_{\sigma}T_{\sigma\sigma}+N_{-\sigma}T_{\sigma,-\sigma}).
	\label{n2}
\end{equation}
The condition of vanishing flow between $L_{1}$ and $T_{A}$ and  
$L_{2}$ and $T_{B}$ are given as
\begin{eqnarray}
	\sum_{\sigma}n_{1\sigma}(\mu_{1}-\mu_{A}) & = & 
	  \sum_{\sigma}(N_{\sigma}-n_{1\sigma})(\mu_{A}-\mu_{2})
  \nonumber \\
	\sum_{\sigma}n_{2\sigma}(\mu_{1}-\mu_{B}) & = & 
	  \sum_{\sigma}(N_{\sigma}-n_{2\sigma})(\mu_{B}-\mu_{2}),
	\label{nocurrent}
\end{eqnarray}
respectively.
This leads to the voltage difference across the sample of
\begin{eqnarray}
	V & = & \frac{1}{e} \sum_{\sigma}(n_{1\sigma}-n_{2\sigma}) 
	(\mu_{1}-\mu_{2})
	\nonumber  \\
	 & = & \frac{V}{2}\frac{1}{N_{+}+N_{-}}\sum_{\sigma} N_{\sigma} 
	 (1+R_{\sigma\sigma}-T_{\sigma\sigma}+R_{-\sigma,\sigma} 
	 -T_{-\sigma,\sigma}).
	\label{muab}
\end{eqnarray}
By use of eq. (\ref{TRresults}) the resistance to the lowest order in 
domain wall scattering is obtained as
\begin{equation}
	R_{\rm w} = \frac{mL}{2e^2\hbar} 
	\frac{1}{\sum_{\sigma}N_{\sigma}k_{F\sigma}} \frac{1}{(1-\zeta^2)^2} 
	[\zeta^2 (1+\zeta\delta)|A_{+}|^2
	 -(\zeta^2+\zeta\delta)|A_{-}|^2],
	\label{R1}
\end{equation}
where $\delta\equiv (N_{+}-N_{-})/ (N_{+}+N_{-})$.
Naively it appears correct to choose 
$N_{\sigma}=\frac{Lm}{\pi\hbar^2 k_{F\sigma}}\equiv D_{\sigma}$, equal to the 
one-dimensional density of state.
This is wrong if the equilibrium situation 
with a steady current flow is considered. 
In fact, the electron with down spin, 
which has a smaller velocity of $k_{F-}$, cannot leave the scattering region 
quickly after the scattering and thus the density
becomes higher by a 
factor proportional to $1/k_{F-}$ (see Fig. \ref{FIGdensity}).
Therefore under a steady flow, the effective density of states 
becomes proportional to $D_{\sigma}/k_{F\sigma}$, namely, 
$N_{\sigma}=D_{\sigma}\frac{(1-\zeta^2)}{(1+\zeta^2)}\frac{1}{(1+\sigma\zeta)}$,
where a prefactor is chosen to conserve the total density, i.e., 
$N_{+}+N_{-}=D_{+}+D_{-}$.
This point has not been discussed in the multi-channel case carried 
out so far\cite{Buttiker85}. 
Then $\delta=-\frac{2\zeta}{1+\zeta^2}$ and we obtain
\begin{equation}
	R_{\rm w} = \frac{h}{e^2} 
	\frac{\zeta^2}{8(1-\zeta^2)} 
	[|A_{+}|^2+|A_{-}|^2].
	\label{Req}
\end{equation}
This result is the same as that obtained based on a linear response 
theory\cite{TZMG99}, where the equilibrium is assumed.

In the case of two-terminals, the system is out of equilibrium and 
the resistance is modified.
The conductance in this case is given as the transmitted current, 
eq. (\ref{Ieq}) with $N_{\sigma}=D_{\sigma}$, divided by the applied voltage, 
$V$.
Thus
\begin{equation}
G=\frac{e^2}{h}\sum_{\sigma} \left(T_{\sigma\sigma} 
+\frac{k_{F,-\sigma}}{k_{F\sigma}}T_{-\sigma,\sigma}\right).
\end{equation}
The wall resistance is written as 
$R_{\rm w}^{\rm (2t)}\equiv G^{-1}-G_{0}^{-1}$, where 
$G_{0}\equiv\frac{2e^2}{h}$ is the conductance without the wall. Thus 
to the lowest order in the wall scattering, 
\begin{equation}
	R_{\rm w}^{\rm (2t)} = \frac{h}{e^2} 
	\frac{\zeta^2}{4(1-\zeta^2)} 
	|A_{+}|^2.
	\label{Rne}
\end{equation}
This resistance contains only backward scattering 
$k_{F\sigma}\rightarrow -k_{F,-\sigma}$.
This result is obtained from the four terminal case, eq. (\ref{R1}), 
by putting $N_{\sigma}=D_{\sigma}$.

Two resistances, (\ref{Req}) and 
(\ref{Rne}) (dashed and solid lines, respectively),  
are plotted in Fig. \ref{FIGR} for the two profiles, 
(\ref{bulkwall}) and (\ref{DWprofiles}). 
In the bulk case, (\ref{bulkwall}), $A_{q}$ is given by
\begin{equation}
	A_{q}=\frac{\pi}{\cosh (\pi q \lambda/2)},
	\label{Abulk}
\end{equation}
and so resistance decays rapidly for $|q|\gtrsim \lambda^{-1}$.
In the case of a linear wall,  (\ref{DWprofiles}), decay is slower (Fig. 
\ref{FIGR});
\begin{equation}
	A_{q}=\pi \frac{\sin q\lambda}{q\lambda}.
	\label{Alinear}
\end{equation}
This difference in the form factor leads to a slightly different 
resistance, but strong suppression of the resistance for 
$k_{F}\lambda\gg1$ is universal.
The difference between the two- and four-terminal cases
is small if the Zeeman splitting is 
large but becomes manifest when $\zeta$ is small. 

Let us mention the case where the leads connected to the sample are 
non-magnetic metals. If we write the hopping probability between the lead 
and the sample as $T^{\rm (L)}_{\sigma}$, the balance of the density 
of states requires
$DT^{\rm (L)}_{\sigma}=N_{\sigma}$, where $D$ is the density of states 
in the lead. Thus the voltage across the sample is determined by the 
same eqs. as (\ref{n1})-(\ref{muab}), and so the resistance is equal to
that for the ferromagnetic lead discussed above.

\section{Summary}
The scattering of the electron by a domain wall is studied theoretically. 
By use of a local gauge transformation, the interaction between the 
electron and the wall is treated perturbatively to the lowest order.
The transmission coefficients are obtained and the resistance was 
calculated using the Laundauer's formula with four-terminals. The 
comparison with the result of linear response theory was made.

\section*{Acknowledgments}
The author thanks H. Matsukawa and Y. Tokura for stimulating discussion.
He also thanks Max Planck Institut fur Mikrostrukturphysik for its 
hospitality during his stay at the early stage of this work and 
Alexander von Humboldt Stiftung for financial support. 
This work is supported by a Grand-in-Aid for Scientific 
Research from the Ministry of Education, Science, 
Sports and Culture.
 

%
%
\end{document}